# Electron irradiation of metal contacts in monolayer MoS$_2$ Field-Effect Transistors

A. Pelella[1,2], O. Kharsah[5], A. Grillo[1,2], F. Urban[1,2,3], M. Passacantando[4], F. Giubileo[2], L. Iemmo[1,2], S. Sleziona[5], E. Pollmann[5], L. Madauß[5], M. Schleberger[5], and A. Di Bartolomeo[1,2,a]

[1] Department of Physics and Interdepartmental Centre NanoMates, University of Salerno, via Giovanni Paolo II, Fisciano, 84084, Italy

[2] CNR-SPIN, via Giovanni Paolo II, Fisciano, 84084, Italy

[3] INFN – Gruppo collegato di Salerno, via Giovanni Paolo II, Fisciano, 84084, Italy

[4] Department of Physical and Chemical Sciences, University of L'Aquila, and CNR-SPIN L'Aquila, via Vetoio, Coppito, L'Aquila, 67100, Italy

[5] Fakultät für Physik and CENIDE, Universität Duisburg-Essen, Lotharstrasse 1, D-47057, Duisburg, Germany

[a] Author to whom correspondence should be addressed. Electronic mail: adibartolomeo@unisa.it.

ABSTRACT

This work deals with the electron beam irradiation of the Schottky metal contacts in monolayer molybdenum disulfide ($MoS_2$) field-effect transistors (FETs). We show that the exposure of the Ti/Au source/drain leads to an electron beam improves the transistor conductance. We simulate the path of the electrons in the device and show that most of the beam energy is absorbed in the metal contacts. Hence, we propose that the transistor current enhancement is due to thermally induced interfacial reactions that lower the contact Schottky barriers. We also show that the electron beam conditioning of contacts is permanent, while the irradiation of the channel can produce transient effects.

INTRODUCTION

Molybdenum disulfide ($MoS_2$) is one of the most studied transition metal dichalcogenides, owing to its layered structure and useful mechanical, chemical, electronic and optoelectronic properties[1–4]. A molybdenum (Mo) atomic plane sandwiched between two sulphur (S) planes constitutes the monolayer that is bonded to other monolayers by weak van der Waals forces to form the bulk material. $MoS_2$ is a semiconductor suitable for several applications[5–9], having 1.2 eV indirect bandgap in the bulk form that widens up to 1.8-1.9 eV and becomes direct in the monolayer[3]. Despite the lower field-effect mobility than graphene[10,11], ranging from few tenths to hundreds[12-15] of $cm^2V^{-1}s^{-1}$, $MoS_2$ field effect transistors (FETs) have recently become very popular as alternatives to graphene FETs[12–17] for next generation electronics based on 2D-materials[18–25].

The fabrication and characterization of devices based on 2D materials greatly rely on the application of electron beam lithography (EBL) or focussed ion beam processing as well as on scanning (SEM) or transmission electron microscopy (TEM), which imply irradiation by charged particles. The exposure to low-energy electrons and/or ions can modify the electronic properties of the 2D materials or their interfaces[9,17,26]. Indeed, structural defects can locally modify the band structure and behave as charge traps, thereby changing the device characteristics both in the case of e-beam[27,28] and ion beam irradiation[29,30]. Conversely, electron beam, ion irradiation or plasma treatments can be intentionally used for nano-incisions[31], pores[32], or to purposely create defects, for instance to reduce the contact resistance[33–35]. Choi et al. reported the effects of 30 keV electron beam (e-beam) irradiation of monolayer $MoS_2$ FETs, showing that irradiation-induced defects act as trap sites reducing the carrier mobility and concentration, and shifting the threshold voltage[36]. A study of point defects in $MoS_2$ using SEM imaging and first-principles calculations, by Zhou et al., demonstrated that vacancies are created by e-beam irradiation at low energies[37], below 30 keV. Durand et al. studied the effects of e-beam on $MoS_2$-based FET reporting an increase of the carriers density and a decrease of the mobility explained as irradiation-induced generation of intrinsic defects in $MoS_2$ and as Coulomb scattering by charges at the $MoS_2/SiO_2$ interface, respectively[38]. Giubileo et al. reported a negative threshold voltage shift and a carrier mobility enhancement under 10 keV electron irradiation of few-layer $MoS_2$ FETs attributed to beam-induced positive charge trapped in the $SiO_2$ gate oxide.

In this paper, we report the spectroscopic and electrical characterization of monolayer $MoS_2$-based FETs, with Schottky Ti/Au contacts, focusing on the effects of low-energy e-beam irradiation. We show that the long exposure of the metal contacts to a 10 keV e-beam in a SEM chamber enhances the transistor´s on-current. We explain such an improvement by radiation-induced lowering of the Schottky barrier at the metal contacts. We perform Monte Carlo simulation to track the e-beam through the device and show that, when the beam is focussed onto the contacts, most of the beam energy is absorbed within the metal. The local heat can induce interfacial reactions that change the chemical composition and structure of the metal/MoS$_2$ interface or can generate or release tensile strain. Both effects cause the lowering of the Schottky barrier and the consequent increase of the transistor current.

Our study shows that electron beam exposure during SEM imaging has non-negligible effects on MoS$_2$ devices; however, it also highlights that a suitable exposure, with the e-beam focused on the contact region, can be conveniently exploited to reduce the contact resistance of the transistor.

EXPERIMENTAL METHODS AND FABRICATION

The MoS$_2$ monolayer flakes were grown via chemical vapour deposition (CVD) in a three-zone split tube furnace, purged with 1000 Ncm³/min of Ar gas for 15 min to minimize the O$_2$ content. The growth SiO$_2$/Si substrate was spin coated with a 1% sodium cholate solution, then a saturated ammonium heptamolybdate (AHM) solution was first annealed at 300 °C under ambient conditions to turn AHM into MoO$_3$ to be used as the source for molybdenum. The target material was placed in a three-zone tube furnace along with 50 mg of S powder, positioned upstream in a separate heating zone. The zones containing the S and AHM were heated to 150 °C and 750 °C, respectively. After 15 min of growth, the process was stopped, and the sample was cooled rapidly.

We realized field-effect transistors using the SiO$_2$/Si substrate (thickness of the dielectric: 285 nm) as the back-gate and evaporating the drain and source electrodes on selected MoS$_2$ flakes through standard photolithography and lift-off processes. The contacts were made of Ti (10 nm) and Au (40 nm) used as adhesion and cover layers, respectively. Figures 1(a) and 1(b) show the SEM top view of a typical device and its schematic layout and measurement setup. The channel is made up from a monolayer flake (as confirmed by Raman and photoluminescence, see below) of width and length of 20 and 4 μm and a nominal thickness of 0.7 nm (typically 1 - 2 nm as measured by AFM).

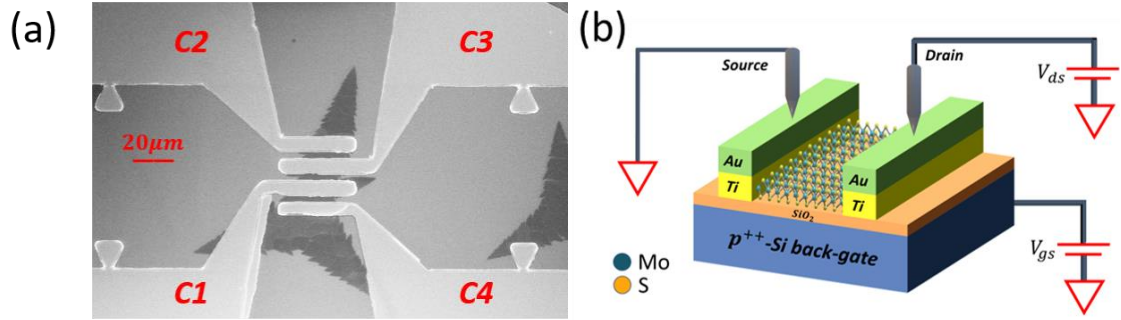

**Figure 1.** (a) SEM image of the $MoS_2$ device and contact labels. (b) $MoS_2$ FET layout and schematic of the common source configuration used for the electrical characterization.

A total of seven $MoS_2$ channels of identically prepared FETs have been characterized by Raman and photoluminescence (PL) spectroscopy just after processing. The measurements were performed with a Renishaw InVia Raman microscope at ICAN (Interdisciplinary Center for Analytics on the Nanoscale). The excitation laser wavelength used was 532 nm and the power density were kept below 0.1 mW/μm² to avoid damage to the $MoS_2$ flakes. Exemplary spectra of the Raman characterization are shown in Figure 2. The chosen reference measurements are spectra obtained from $MoS_2$ flakes on the same substrates, which were also in contact with photoresist and various solvents during the processing and lift-off for the production of the FETs, but are not in contact with metal contacts themselves. As can be seen in Figure 2, the shape of the PL spectra (a) and the difference of the Raman modes (b) differ significantly. The PL intensity (sum of all excitons and trions) for non-contacted $MoS_2$ flakes is higher by a factor of 1.7 ± 0.8 than for contacted $MoS_2$. The mode difference for non-contacted and contacted $MoS_2$ is 21.3 ± 0.7 cm$^{-1}$ and 19.7 ± 0.7 cm$^{-1}$, respectively. Both the changes in PL and Raman mode difference can be associated with built-in strain and changes in the electronic properties and band structure of the MoS$_2$ sheets[39–43]. When comparing contacted MoS$_2$ with non-contacted MoS$_2$ we find a reduction of tensile strain by (0.46 ± 0.28) % and an increase in the electron doping of 0.44 ± 0.36 × 10$^{13}$ electrons per cm² for the contacted 2D material. At this point we therefore conclude that a metal contact significantly alters the properties of monolayer MoS$_2$.

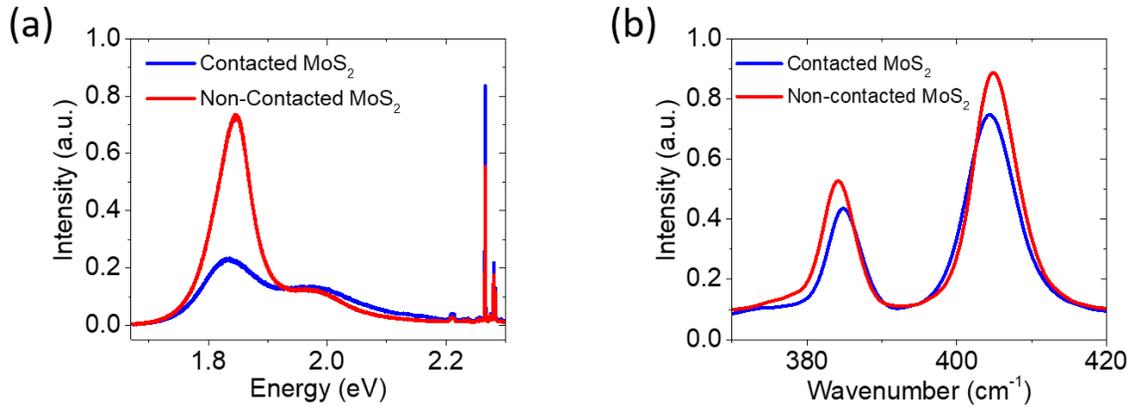

**Figure 2.** a) Photoluminescence and b) Raman spectrum of monolayer MoS$_2$ after FET processing. Blue: contacted MoS$_2$ monolayer flake, red: non-contacted monolayer MoS$_2$ flake.

In the following, most of the electrical characterization refers to the transistor between the contacts labelled as C2 and C3 in Figure 1(a). The contact C3 was used as the drain and C2 as the grounded source. The electrical measurements were carried out inside a SEM chamber (LEO 1530, Zeiss), endowed with two metallic probes with nanometer positioning capability, connected to a Keithley 4200 SCS (source measurement units, Tektronix Inc.), at room temperature and pressure of about 10$^{-6}$ mbar. The e-beam of the SEM, set to 10 keV and 10 pA, was used for the time-controlled irradiation of specific parts of the device.

RESULTS AND DISCUSSION

The output ($I_{ds}$-$V_{ds}$) and the transfer ($I_{ds}$-$V_{gs}$) characteristics of the transistor are shown in Figure 3(a) and 3(b), respectively. The output curve shows rectification with the forward current appearing at negative $V_{ds}$, typical of a p-type Schottky diode, while the transfer characteristic shows an n-type transistor. This apparently contradictory behaviour has been previously reported for MoS$_2$ and WSe$_2$ transistors and explained by the formation of two back-to-back and possibly asymmetric Schottky barriers at the contacts[44,45]. The forward current at negative $V_{ds}$ is caused by the image force barrier lowering of the forced junction (that is the drain, C3, in our case), while the reverse current at $V_{ds}$>0 V is limited by the grounded junction at the source (C2) contact. As the barrier lowering is more effective on the forced junction, being the voltage directly applied to it, the negative bias gives rises to the higher (apparently forward) current.

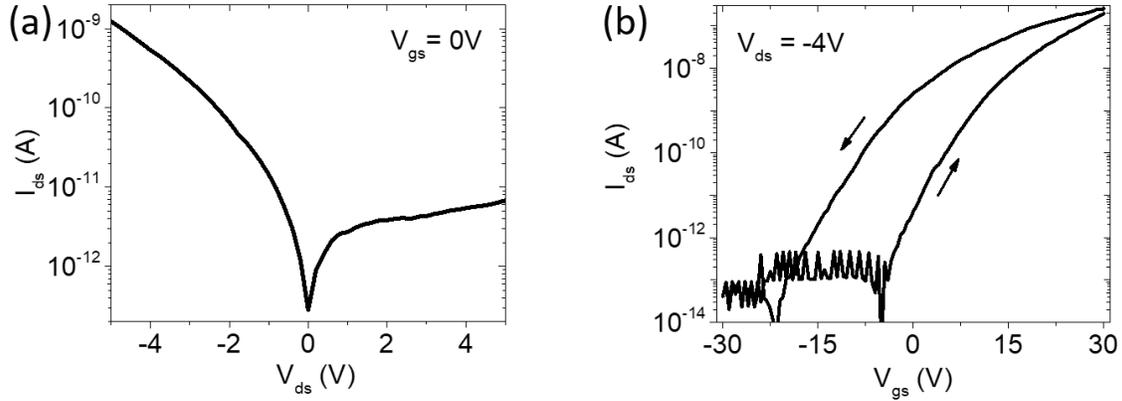

**Figure 3.** Output (a) and transfer (b) characteristics of the device between C2 and C3 contacts, with C3 used as the drain and C2 as the grounded source.

After the initial electrical characterization, we performed two sets of exposures to the SEM electron beam. Each exposure lasted 300 s corresponding to a fluence of ~180 $\frac{e^-}{nm^2}$, over a surface of ~100 μm². The two sets of irradiations were performed first on the drain contact (C3) and then on the grounded source contact (C2). A final exposure of the MoS$_2$ channel to the e-beam was performed as well.

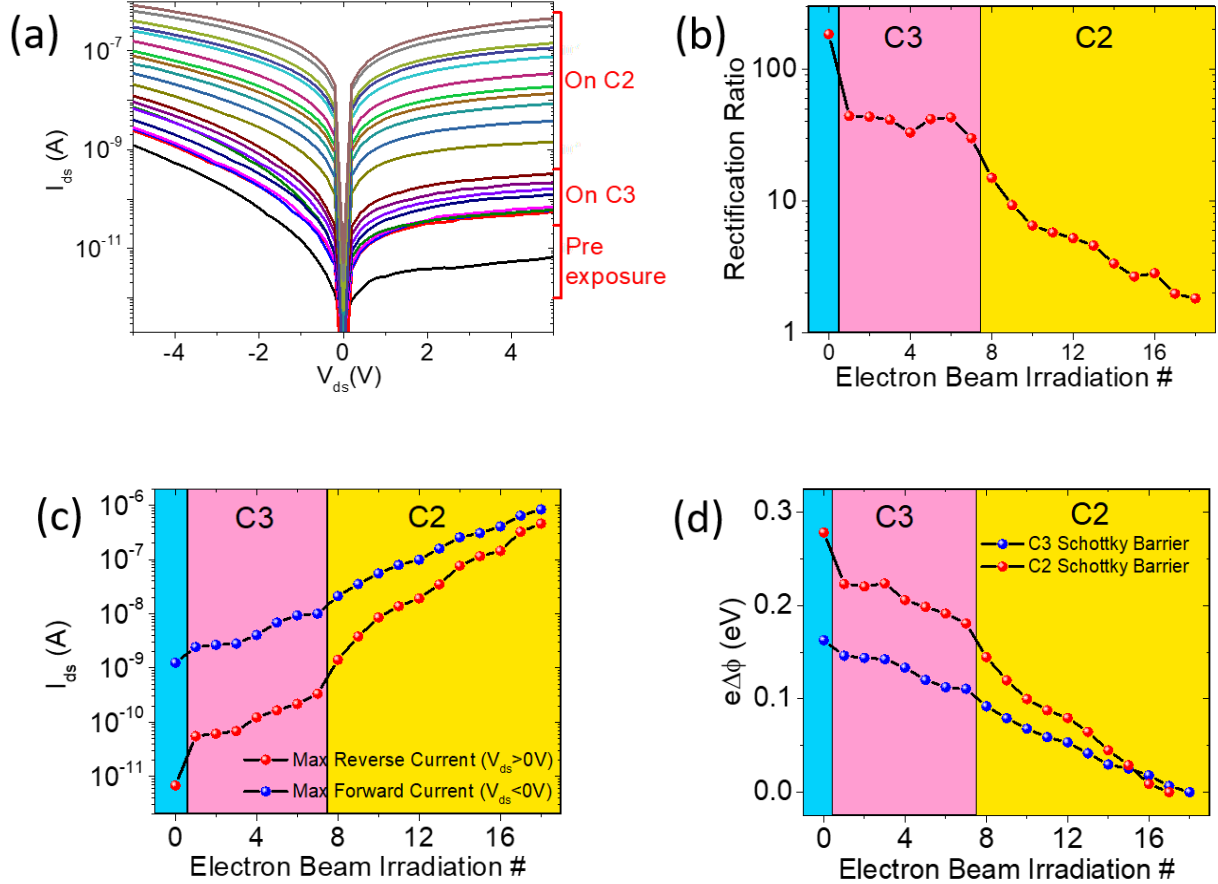

**Figure 4.** (a) Output characteristics of the transistor between contacts C2-C3 exposed to two sets of electron irradiations performed first on contact C3, then on C2. (b) Rectification ratio and (c) maximum forward and reverse current, at $V_{ds}=\pm 5$ V, as a function of the irradiation number. (d) Calculated Schottky barriers at the contact C2 and C3 as a function of the irradiation number.

Figure 4 summarizes the obtained results. Starting from the bottom (black) line in Figure 4(a) representing the output curve of the unexposed device, the current increases with the e-beam exposures. We note two major discontinuities in the sequence of $I_{ds}$-$V_{ds}$ curves, corresponding to the start of the two irradiations sets. These gaps are likely due to the uncontrolled exposure of the whole device during the selection of the drain (C3) and grounded source (C2) contact area for the respective irradiation sets.

A different behavior of the forward with respect to the reverse current can be observed in Figure 4(a) and a distinction of the effects of the irradiations on the drain (C3) and the grounded source (C2) can be made. While the irradiation of the drain increases both the forward and the reverse currents, keeping the rectification ratio almost constant (see Figure 4 (b)), the irradiation of the source augments the reverse current in a faster way, rendering the output curves more symmetric. Figure 4(b) shows that repeated irradiations on the drain contact (C3) do not change the rectification ratio (at $V_{ds}=\pm 5$ V) while the irradiation set on the grounded source contact (C2) dramatically decreases the rectification ratio. Figure 4(c) shows that the maximum reverse and forward currents, at $V_{ds}=\pm 5$ V, have different variation rates when the irradiation is either on drain or source. Noticeably, Figure 4(c) shows that the increase of both the reverse and forward current is an exponential function of the fluence, that is proportional to and can be parametrized by the irradiation number.

As the shape and the current intensity of the output characteristics is related to the Schottky barrier heights at the contacts, the exponentially increasing current and the changing rectification ratio point to radiation-induced Schottky barrier lowering. The energy release in the metal contacts can modify the chemistry of the metal/$MoS_2$ interface or create stress and defects that can lead to a lowering of the barrier and a consequent contact resistance reduction. We note that the reduction of the contact resistance by chemical reactions between the metal contacts and $MoS_2$ channel has been reported for electron-beam metal deposition[46] and contact laser annealing[47]. A disordered, compositionally graded layer, composed of Mo and $Ti_xS_y$ species forms at the surface of the $MoS_2$ crystal following the deposition of Ti, and thermal annealing in the 100−400 °C temperature range can cause further chemical and structural changes at the Ti/MoS$_2$ interface[48,49]. Similarly, tensile strain has been demonstrated to induce considerable Schottky and tunnelling barrier lowering[50].

As the forward current at $V_{ds}<0$ V is limited by the Schottky barrier at the drain contact (C3), while the reverse current at $V_{ds}>0$ V is limited by the Schottky barrier at the grounded source contact C2 (which are the reverse-biased junctions for negative and positive $V_{ds}$, respectively), the output curves of Figure 4(a), that correspond always to reverse current, can be used to extract the behaviour of the Schottky barriers as a function of the fluence (i.e. the e-beam irradiation number). Let us consider the thermionic current through a reverse biased Schottky barrier[51]:

$$I_n = I_{sn}\left[e^{\frac{eV_a}{nkT}}-1\right] = \left[SA^*T^2 e^{-\frac{e\varphi_{Bn}}{kT}}\right]\left[e^{\frac{eV_a}{nkT}}-1\right] \approx -SA^*T^2 e^{-\frac{e\varphi_{Bn}}{kT}} \qquad (1)$$

where $\varphi_{Bn}$ and $I_{sn}$ are the barrier height and the reverse saturation current at the n-th e-beam irradiation, S is the junction area and A* is the Richardson constant, k is the Boltzmann constant, T is the temperature, n is the ideality factor, and $V_a$ is the negative voltage across the barrier that makes $e^{\frac{eV_a}{kT}}\approx 0$. Let us define $I_0$ as the reverse saturation current associated to the last e-beam exposure, i.e. to the minimum barrier height $\varphi_{B0}$. Then, Equation (1) can be used to evaluate the variation of the Schottky barrier, $\Delta\varphi_{Bn}=\varphi_{Bn}-\varphi_{B0}$, as a function of the irradiation number:

$$\ln\left(\frac{I_n}{I_0}\right) = -\frac{e\Delta\varphi_{Bn}}{kT} \rightarrow \Delta\varphi_{Bn} = -\frac{kT}{e}\ln\left(\frac{I_n}{I_0}\right) \qquad (2)$$

The barrier variation, $\Delta\varphi_{Bn}$, obtained taking $I_n$ and $I_0$ at $V_{ds}=\pm 5$ V, is shown in Figure 4(d) for both source (C2) and drain (C3) contacts. The plot indicates that the two barriers behave differently for the irradiation of C2 or C3. While the beam irradiation of either contact results in a lowering of both Schottky barrier, the barrier decrease is faster for the irradiation of the grounded source.

Besides, the Schottky barrier at the source contact is the most affected by the irradiation of the source.

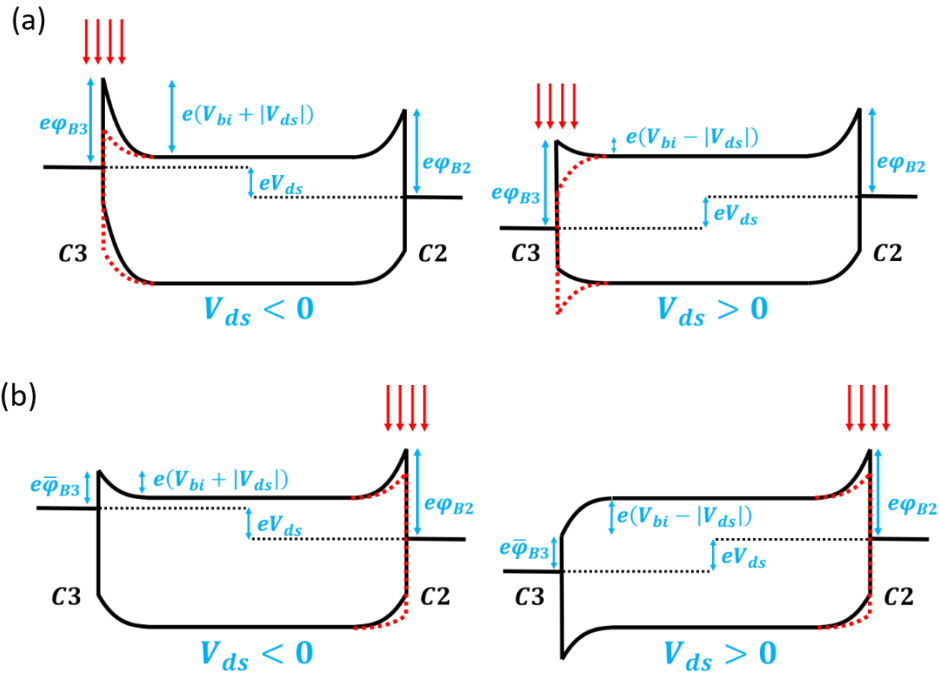

**Figure 5.** Low-bias energy band diagrams (black) and their modification under electron irradiation (red) of C3 (a) and of C2 (b) contact resulting in barrier lowering ($\bar{\varphi}_B$).

To explain these results, we propose the model based on the energy band diagrams shown in Figure 5. A negative (positive) voltage applied to the drain contact (C3) causes an upward (downward) shift of the energy bands in the drain region. Electron beam irradiation of the contact lowers the Schottky barrier and the relative built-in potential, as shown by the red dashed lines in Figure 5. The reduction of a Schottky barrier and of its associated built-in potential, at the irradiated contact, results also in the lowering of the unexposed barrier, which can experience a stronger potential drop due to the reduced contact resistance of the first contact. Figure 5(a) represents the situation in which the e-beam is focussed on the biased drain contact (C3). At $V_{ds} < 0$ V, the current is limited mainly by the drain contact barrier which is lowered by the successive irradiations causing the exponential increase of the maximum forward current. At $V_{ds} > 0$ V, the current is limited by the unirradiated source contact (C2) barrier, and its dependence on the irradiation cycle is caused by the lowering of the built-in potential at the drain (C3). As the barrier and built-in lowering is the same, the rectification ratio remains about constant. For irradiation of the grounded source (C2, Figure 5(b)), the current increases due to a similar mechanism, with the difference that the drain contact barrier limits the current for $V_{ds} > 0$ V to a lesser extent, having been already irradiation-lowered. Therefore, the reverse current increases faster with the repeated irradiation and the rectification ratio decreases.

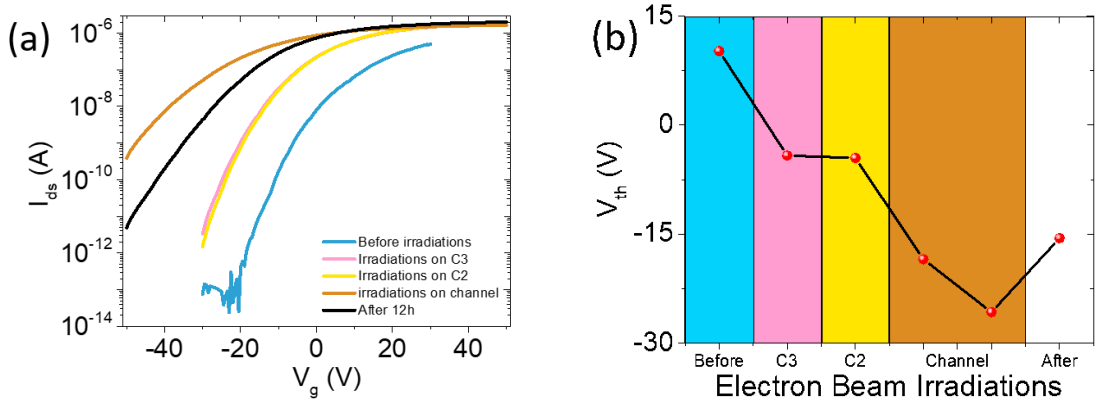

**Figure 6.** (a) FET transfer characteristics before and after e-beam irradiations of contacts C3 and C2 as well as of channel. (b) Left shift of the threshold voltage extrapolated from the transfer characteristics over the e-beam exposure.

The effect of the irradiation on the transfer characteristic of the transistor is shown in Figure 6 and confirms the radiation-induced increase of the channel current. Besides, Figure 6(a) shows that the e-beam, independently of onto which contact it is focused on, causes a left shift of the transfer curve. Such a shift corresponds to a decrease of threshold voltage, defined as the x-axis intercept of the linear fit of the transfer curve on linear scale. The threshold voltage as a function of the irradiation is displayed in Figure 6(b). While the e-beam exposure of the contacts provokes a left-shift (the transfer curves are taken at the end of the two irradiation sets on the drain (C3) and grounded source (C2), respectively), further left-shift of the threshold voltage is observed when two successive irradiations are performed on the channel region.

The observed negative shift of the threshold voltage has been reported and discussed before[27]. It can be explained by the pile up of positive charge in trap states of the $SiO_2$ gate dielectric. The e-beam exposures produce electron−hole pairs in the $SiO_2$ gate oxide: while mobile electrons are easily swept by the applied bias, the positive charges can be trapped for long times[27]. The positive charge storage acts as an extra gate (similarly to the gating effect under light irradiation[52,53]) and enhances the n-type doping of the channel.

Indeed, Figure 6 shows that there is a slight recovery of the threshold voltage after 12 h annealing at room temperature. However, we highlight that, as demonstrated by Figure 6(a), the maximum channel current, which is limited by the contact resistances, remains unchanged after the annealing, demonstrating that the irradiation-induced improvement of the contacts is permanent.

To further confirm our model, we performed a Monte Carlo simulation to track the path of the electrons under the contacts and in the channel region (Figure 7(a-b)), using the CASINO software package[54–56]. We simulated a 10 keV beam with one million electrons and a radius beam of 10 nm. The cathodoluminescence spectrum (Figure 7(c)) shows that electrons lose their energy and are stopped (Figure 7(d)) mostly in the Ti/Au metal stack, while they reach and are absorbed in the Si substrate when the irradiation is on the channel. The high release of energy in the metal contacts, similarly to thermal annealing[57,58], induces Ti/$MoS_2$ reactions and creates contact with reduced

Schottky barrier and contact resistance. Conversely, when we directly irradiate the MoS$_2$ channel, energy is prevalently adsorbed in the Si bulk, and its effect manifest only through the positive charge traps generated in the SiO$_2$ layer.

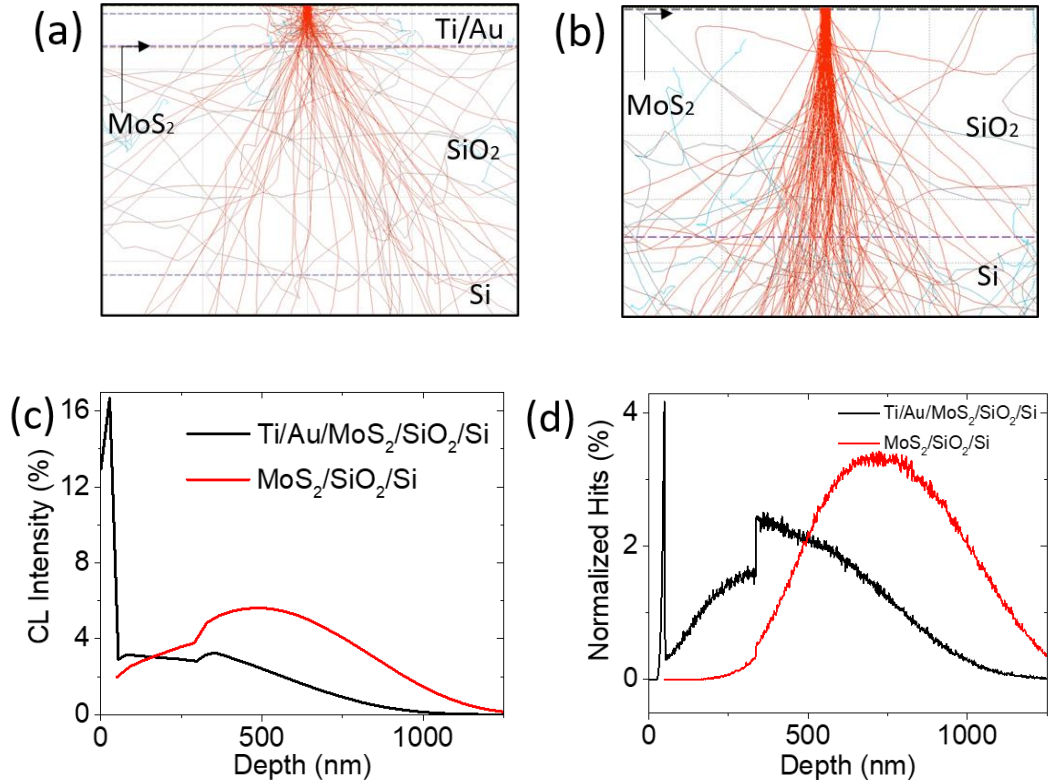

**Figure 7.** Monte Carlo simulation using CASINO v2 of e-beam irradiation of the device (a) contacts and (b) of the MoS$_2$ channel. (c) Simulated cathodoluminescence intensity through the sample, with the e-beam focused onto contacts and onto the flake. (d) Simulation of the electron´s penetration depth through the sample.

CONCLUSIONS

We investigated the effects of 10 keV electron beam irradiation of the Schottky metal contacts in MoS$_2$ based FETs. Spectroscopic analysis by Raman and photoluminescence shows that the presence of metal contacts changes the properties of monolayer MoS$_2$ with respect to strain and doping. The electrical measurements revealed that electron beam irradiation improves the device conductance, reduces the rectification of the output characteristic and causes a left-shift of the threshold voltage. To explain such a feature, we propose that the energy absorbed in the metal contacts induces interfacial reactions that lower the Schottky barrier at the contacts and improve the contact resistance. We corroborate our model by direct measurement of the Schottky barrier height variation and by simulation of the electron trajectories in the contact regions.

ACKNOWLEDGEMENTS

The authors thank MIUR, projects Pico & Pro ARS01_01061 and RINASCIMENTO ARS01_01088, and German Research Foundation (DFG), project number 406129719, for the financial support. Raman and PL spectroscopy was performed at the ICAN, a core facility founded by the German Research Foundation (DFG, reference RI_00313).